\documentstyle[12pt]{article}
\newtheorem{lemma}{Proposition}

\begin{document}
\title{
Construction of trigonometric Toda $r$-matrices via Hamiltonian reduction
of the cotangent bundle over loop groups.
}
\author{G.E.Arutyunov
\thanks{Steklov Mathematical Institute,
Vavilov 42, GSP-1, 117966, Moscow, Russia; arut@class.mian.su}
}
\date {}
\maketitle
\begin{abstract}
By applying the Hamiltonian reduction method to the cotangent bundle over
loop groups we recover the well-known
classical trigonometric $r$-matrices of the periodic Toda lattice.
\end{abstract}
\newpage
\section{Introduction}
The aim of this paper is to show that the
trigonometric $r$-matrices of the Toda model \cite{BD,J} can be derived by
applying the Hamiltonian reduction method to some infinite-dimensional
phase space.

It is known that finite-dimensional dynamical systems of Toda- or
Calodgero-types can be obtained by the Hamiltonian
reduction of geodesic motions on the cotangent bundles of
semi-simple Lie groups \cite{OP,FO}. In doing so,
$L$-operator, coming in the Lax representation, arises as a point
on the reduced phase space ${\cal P}$, whereas the Lax representation
$\frac{dL}{dt}=[M,L]$ -  as the equation of motion on ${\cal P}$.
Besides this construction gives the $L$-operator a geometric meaning,
it also allows one to calculate the classical $r$-matrix
and to prove thereby under some additional assumptions
the kinematic integrability of a system.

In recent paper \cite{ABT} a nice computational
scheme along the lines described above was worked out in detail
and applied to recover the known \cite{AT} dynamical $r$-matrices of the
Calodgero-Moser-Sutherland models.
Note that $L$-operators in question
do not contain the spectral parameter that is needed to construct
the action-angle variables. Thus, we come across a question to develop
a Hamiltonian reduction scheme that provides the Lax representation
of a dynamical system with the spectral parameter.

This paper is the straightforward generalization of the construction of
\cite{ABT} to the affine case. By considering the periodic Toda model as
an example, we show that its Lax representation with the spectral
parameter can be derived by applying the Hamiltonian reduction to the
cotangent bundle over a (twisted) loop group corresponding to one of
simple Lie groups. In this construction the loop parameter is naturally
identified with the spectral one. The main point of our reduction
is that the reduced phase space turns out to be finite-dimensional.
$r$-matrices are calculated in the same manner as in \cite{ABT} and
they are nothing but the trigonometric solutions of the classical
Yang-Baxter equation.

Finally note that the relation between trigonometric $r$-matrices
and affine algebras is well known (see, for example, \cite{FT}).
As it will be clear from our analysis, from the Hamiltonian
reduction point of view, this relation is due to the existence of
the Iwasawa decomposition for an infinite-dimensional loop group.
For the sake of simplicity we confine ourselves with consideration
of affine algebras of the height 1.

\section{Affine algebras and Iwasawa decomposition}
Denote by $\hat{{\cal G}}$ an affine algebra of rank $l$.
The factor-algebra ${\cal G}=\hat{{\cal G}}/{\bf
C}c$ of $\hat{{\cal G}}$ by its one-dimensional center $c$ has
two basic realizations: homogeneous and principal that arise from the
corresponding realizations of $\hat{{\cal G}}$ \cite{K}.

Let $A_{ij}$ be the Cartan matrix of $\hat{{\cal G}}$ and $a_i$ be the
labels of the Dynkin diagram of the root system. Recall that as an
abstract algebra ${\cal G}$ is generated by $3(l+1)$ generators
$\{e_i,h_i,f_i\}$, $i=0,\ldots,l$ and the following relations
\begin{equation}
\begin{array}{ll} [h_i,h_j]=0,& [e_i,f_j]=\delta_{ij}h_i,            \\
\mbox{$[$} h_{i},e_{j}\mbox{$]$}= A_{ij}e_{j}, &
\mbox{$[$} h_{i},f_{j}\mbox{$]$}=-A_{ij}f_{j},
\end{array}
\label{abs}
\end{equation}
$$(\mbox{ad}e_i)^{1-A_{ij}}e_j=(\mbox{ad}f_i)^{1-A_{ij}}f_j=0,$$
and one additional relation $\sum_{i=0}^{l}a_ih_i=0$ that fixes
the central element $c=\sum_{i=0}^{l}a_ih_i$ of (\ref{abs}) equal to
zero. ${\cal G}$ is a graded Lie algebra.

1. Homogeneous gradation of $\cal G$.\\
Put $\deg e_i=\deg f_i=\deg h_i=0,~i=1,\ldots,l$ and $\deg e_0=-\deg
f_0=1$, where the triple $(e_0,h_0,f_0)$ corresponds to the special vertex
of the Dynkin diagram. It can be easily shown that being graded in
such a way $\cal G$ is isomorphic to the loop algebra
$\bar{{\cal G}}[\lambda,\lambda^{-1}]$ over a simple finite-dimensional
Lie algebra $\bar{{\cal G}}$ that is obtained from $\hat{{\cal G}}$
by deleting the special vertex of the Dynkin diagram:
${\cal G}=\sum_{k\in{\bf Z}}\bar{{\cal G}}\otimes \lambda^k.$
Note that as a linear space  ${\cal G}^k=\bar{{\cal G}}\otimes \lambda^k$
is generated by
\begin{equation}
E_{\alpha}^k=e_{\alpha}\otimes
\lambda^k,~~~ h_i^k=h_i\otimes \lambda^k, ~~~i=1,\ldots l, \label{lb1}
\end{equation}
where $\alpha\in \Delta$, $\Delta$ is the set of roots of $\bar{{\cal G}}$.

2. Principal gradation of $\cal G$.\\
The principal gradation ${\cal G}=\sum_{k\in{\bf Z}}{\cal G}^k$ is
assigned by setting $\deg e_i=-\deg f_i=1,~i=0,\ldots,l$ and $\deg h_i=0$.
It can be proved that in this gradation $\cal G$ is isomorphic
to the algebra of twisted loops
${\cal G}=\sum_{k\in{\bf Z}}\bar{{\cal G}}_{k\bmod h}\otimes \lambda^k$
constructed with the help of the decomposition of $\bar{{\cal G}}$
in eigenspaces of a Coxeter automorphism $C$:
$$
\bar{{\cal G}}_k=\{x\in\bar{{\cal G}}, C(x)=\omega^k x,~~
\omega=e^{\frac{2\pi i}{h}}\},
$$
where $h$ is the Coxeter number for $\hat{{\cal G}}$.

Now let us choose in ${\cal G}$ the linear basis of homogeneous elements
under the principal gradation.
If $k\neq 0\bmod h$, then the space ${\cal G}^k$ is linearly
generated by the elements $E_{\alpha}^k=e_{\alpha}\otimes \lambda^k$, where
$\alpha\in \Delta^{k}$ and $\Delta^{k}=\{\alpha\in \Delta,~~e_{\alpha}\in
\bar{{\cal G}}_k\}$, i.e. $e_{\alpha}$ runs the set of all root vectors
from  $\bar{{\cal G}}_k$. If $k= 0\bmod h$, then
vectors $h_i^k=h_i\otimes \lambda^k$ form a basis in ${\cal G}^k$.
Therefore,
$$
{\cal G}=\sum_{k\neq 0\bmod h}\sum_{\alpha\in \Delta^{k\bmod h}}{\bf C}
E_{\alpha}^{k}+\sum_{k}{\bf C}h_{i}^{kh}.
$$

It is useful to introduce the following notation. Denote by
$n_{+}(n_-)$ a subalgebra in $\cal G$ generated as a vector space by
multiple commutators of elements $e_i$ $(f_i)$, $0\leq i\leq
l$ and denote by $\bar{h}$ the Cartan subalgebra in $\bar{{\cal G}}$.

\begin{lemma}
In the homogeneous gradation of ${\cal G}$ one has
\begin{equation}
\begin{array}{l}
n_+=(\bar{n}_+\otimes 1)\oplus (\bar{{\cal G}}\otimes \lambda{\bf C}
[\lambda]),\\
n_-=(\bar{n}_-\otimes 1)\oplus (\bar{{\cal G}}\otimes
\lambda^{-1}{\bf C}[\lambda^{-1}]),
\end{array}
\label{nil}
\end{equation}
where $\bar{n}_{\pm}$ are nilpotent subalgebras in the triangular
decomposition
$\bar{{\cal G}}=\bar{n}_{+}\oplus\bar{h}\oplus\bar{n}_{-}$.
\end{lemma}
Proof. Chevalley generators (\ref{abs}) of ${\cal G}$ can
be defined in the following way. Choose in $\bar{\cal G}$ a vector
$\bar{e}_{0}$ ($\bar{f}_{0}$) corresponding to the highest (lowest) root and
normalize it by the condition $(\bar{e}_{0},\bar{f}_{0})=1$. Denote
by $\bar{e}_i,(\bar{f}_i)$, $1\leq i\leq l$ the root vectors corresponding
to simple positive (negative) roots and commuting as
$[\bar{e}_i,\bar{f}_i]=\bar{h}_i$. Then Chevalley generators for
${\cal G}$ are given by
$$
e_{0}=\bar{f}_{0}\otimes \lambda,~~~
f_{0}=\bar{e}_{0}\otimes \lambda^{-1},~~~
h_{0}=[e_{0},f_{0}]
$$
together with $\{\bar{e}_i,\bar{h}_i,\bar{f}_i\}_{1\leq i\leq l}$.
Now considering multiple commutators of generators $e_i$ we get
the statement. Realization of Chevalley generators used in Proposition
1 is called homogeneous.

Let $\cal G$ has the homogeneous gradation. Denote by
${\cal G}^R$ the real normal form of $\cal G$. ${\cal G}^R$
is generated by the linear basis
$E_{\alpha}^k=e_{\alpha}\otimes \lambda^k,~h_i^k=h_i\otimes \lambda^k$
over $\bf R$. Consider in ${\cal G}^R$
an involution $\tau$ defined by the following formulas
\begin{equation}
\tau (E_{\alpha}^k)=-E_{-\alpha}^{-k},~~~\tau (h_i^k)=-h_i^{-k}.
\label{inv}
\end{equation}

\begin{lemma}
An algebra $\cal K$ of $\tau$-stable points in ${\cal G}^{R}$
is generated by a linear basis
\begin{equation}
{\cal K}=\sum_{\stackrel{\alpha\in \Delta}{k>0}}{\bf R}V_{\alpha}^{k}
+\sum_{\alpha\in \Delta_+} {\bf R}V_{\alpha}^{0}+
\sum_{\stackrel{1\leq i\leq l}{k>0}} {\bf R} H_i^k,
\label{lb}
\end{equation}
where $V_{\alpha}^k=E_{\alpha}^k-E_{-\alpha}^{-k}$ and $H_i^k=h_i^k-h_i^{-k}$.
\end{lemma}
The proof is trivial.

Recall that ${\cal G}^R$ is equipped with a nondegenerate symmetric
bilinear form invariant with respect to the adjoint action of ${\cal G}^R$.
It is easy to see that this form is negatively defined on
$\cal K$, i.e. $\cal K$ is a compact subalgebra in
${\cal G}^R$. Thus, ${\cal G}^R$ is a direct sum
${\cal G}^R={\cal K}\oplus{\cal M}$ of algebra $\cal K$
and a vector space  $\cal M$:
\begin{equation}
{\cal M}=\sum_{\stackrel{\alpha\in \Delta}{k>0}}{\bf R}W_{\alpha}^{k}
+\sum_{\alpha\in \Delta_+} {\bf R}W_{\alpha}^{0}+
\sum_{\stackrel{1\leq i\leq l}{k\geq 0}} {\bf R} T_i^k,
\label{lbv}
\end{equation}
where $W_{\alpha}^k=E_{\alpha}^k+E_{-\alpha}^{-k}$ and $T_i^k=h_i^k+h_i^{-k}$.
This decomposition is an analog of the Cartan decomposition
for a simple finite-dimensional Lie algebra over ${\bf R}$.

In what follows an important role will be played by an analog of the
Iwasawa decomposition. Note that we can
decompose any $x\in {\cal G}$ over the homogeneous basis defined above:
$$
x=
\sum_{\stackrel{\alpha\in \Delta}{k>0}}
a_{\alpha}^{k}(E_{\alpha}^k-E_{-\alpha}^{-k}) +
\sum_{\alpha\in \Delta_+}
a_{\alpha}^{0}(E_{\alpha}^0-E_{-\alpha}^{0})+
\sum_{\stackrel{1\leq i\leq l}{k>0}} a_i^k(h_i^k-h_i^{-k})+
$$
$$
\sum_{\stackrel{\alpha\in \Delta}{k>0}}
b_{\alpha}^{k}(E_{\alpha}^k+E_{-\alpha}^{-k}) +
\sum_{\alpha\in \Delta_+}
b_{\alpha}^{0}(E_{\alpha}^0+E_{-\alpha}^{0})+
\sum_{\stackrel{1\leq i\leq l}{k\geq 0}} b_i^k(h_i^k+h_i^{-k}).
$$
This decomposition can be rearranged in the following way
$$
x=
\sum_{\stackrel{\alpha\in \Delta}{k>0}}
(a_{\alpha}^{k}-b_{\alpha}^{k})V_{\alpha}^k +
\sum_{\alpha\in \Delta_+}
(a_{\alpha}^{0}-b_{\alpha}^{0})V_{\alpha}^0+
\sum_{\stackrel{1\leq i\leq l}{k>0}} (a_i^k-b_i^k)H_i^k+
$$
$$
\sum_{\stackrel{\alpha\in \Delta}{k>0}}
2b_{\alpha}^{k}E_{\alpha}^k +
\sum_{\alpha\in \Delta_+}
2b_{\alpha}^{0}E_{\alpha}^0+
\sum_{\stackrel{1\leq i\leq l}{k>0}} 2b_i^k h_i^k+
\sum_{1\leq i\leq l} 2b_i^0 h_i.
$$
By virtue of Proposition 1, an element
$\sum_{\stackrel{\alpha\in \Delta}{k>0}}
2b_{\alpha}^{k}E_{\alpha}^k +
\sum_{\alpha\in \Delta_+}
2b_{\alpha}^{0}E_{\alpha}^0+
\sum_{\stackrel{1\leq i\leq l}{k>0}} 2b_i^k h_i^k
$
is in $n_+$. Therefore, ${\cal G}^{R}$ decomposes in direct sum of three
subalgebras: ${\cal G}^{R}={\cal K}\oplus n_{+}\oplus a$, where $a$ is an
embedding of $\bar{h}$ in ${\cal G}^R$.
We refer to this decomposition as to Iwasawa one for
${\cal G}^R$. The Iwasawa decomposition has a global analog
$G=N_+A K$, where $G$ denotes the loop group and
$N_+,~A$ and $K$ are simply connected Lie groups with Lie algebras
$n_{+},~ a$ and ${\cal K}$ respectively.
Note that $N_+$
can be treated as the group consisting of boundary values
of holomorphic mappings
$$ \gamma:~~ \{z:~|z|<1\}\rightarrow \bar{G}, $$
such that $\gamma(0)$ takes a value in a nilpotent subgroup of a group
$\bar{G}$ corresponding to positive roots \cite{PS}.

Now consider the Cartan decomposition for the principal realization of
${\cal G}$.  Since the involution is given by the same formula
(\ref{inv}) the eigenspaces of  $\tau$ in
${\cal G}=\sum_{k\in{\bf Z}}\bar{{\cal G}}_{k(\bmod h)}\otimes \lambda^k$
have the form
\begin{equation}
{\cal K}=\sum_{\stackrel{k>0}{k\neq 0 \bmod h}}\sum_{\alpha\in \Delta^{k
\bmod h}} {\bf R}V_{\alpha}^{k}+\sum_{\stackrel{k>0}{1\leq i\leq l}}
{\bf R}H_i^{kh}
\label{com}
\end{equation}
and
\begin{equation}
{\cal M}=\sum_{\stackrel{k>0}{k\neq 0 \bmod h}}\sum_{\alpha\in \Delta^{k
\bmod h}} {\bf R}W_{\alpha}^{k}+\sum_{\stackrel{k\geq 0}{1\leq i\leq l}}
{\bf R}T_i^{kh}.
\label{vec}
\end{equation}
{}From this representations for
${\cal K}$ and ${\cal M}$ one can immediately derive
the Iwasawa decomposition for the loop algebra twisted by the Coxeter
automorphism.

\section{Hamiltonian reduction of the cotangent bundle over Lie group}
Let $G$ be a finite-dimensional Lie group with a Lie algebra $\cal G$ and
$T^{*}G$ be a cotangent bundle over $G$. By using left group translations
we identify $T^{*}G$  with $G\times {\cal G}^*$, where ${\cal G}^*$ is
a dual to $\cal G$. Thus, a point in $T^{*}G$ is labeled by a pair
$(g,\xi)$, $g\in G$ and $\xi \in {\cal G}^*$.
The canonical nondegenerate Poisson structure on $T^{*}G$ is
\begin{equation}
\{\phi,\psi\}=\{\phi,\psi\}_{\mbox{Kir}}+\partial^i
\phi L_i \psi- \partial^i \psi L_i \phi,~~~\phi,\psi\in
\mbox{Fun}(T^{*}G).
\label{pbr}
\end{equation}
Recall the notation used in the last formula.
Choose in $\cal G$, ${\cal G}^*$
dual bases $\{e_i\}$  $\{e^i\}$, i.e. $<e^i,e_j>=\delta_i^j$
and $\xi=\xi_i e^i$. The bracket
$\{\phi,\psi\}_{\mbox{Kir}}$ is the Kirillov bracket of functions
$\phi$ and $\psi$ restricted on ${\cal G}^*$:
\begin{equation}
\{\phi,\psi\}_{\mbox{Kir}}=f_{ij}^k \xi_k\partial^i \phi \partial^j \psi,
\label{cbr}
\end{equation}
where $f_{ij}^k$ are the structure constants of $\cal G$ in a basis $\{e_i\}$
and
$\partial^i \phi=\frac{\partial}{\partial\xi_i}\phi$. Operator $L_i$
is treated as the left invariant vector field on $G$ generated by $e_i$:
$$
(L_i\phi)(g)=\frac{d}{dt}|_{t=0}\phi(ge^{te_i}).
$$
Group $G$ acts on $T^{*}G$ by left and right translations:
\begin{equation}
g'\cdot (g,\xi)=(g'g,\xi),~~~
(g,\xi)\cdot g'=(g(g')^{-1},\mbox{Ad}^*_{g'}\xi).
\end{equation}
These actions generate fundamental vectors fields on $T^{*}G$
being hamiltonian. The corresponding hamiltonians have the form:
\begin{equation}
H^{L}_X(g,\xi)=<\xi,g^{-1}Xg>,~~~H^{R}_X(g,\xi)=-<\xi,X>,~~X\in{\cal G}.
\label{gam}
\end{equation}
As the matter of fact from (\ref{pbr}) we get
$$
\{H_{X}^{R},\psi\}=<\mbox{ad}_{X}^{*}\xi,\nabla \psi>-L_X\psi=
\frac{d}{dt}|_{t=0}\psi(ge^{-tX},\mbox{Ad}^*_{e^{tX}}\xi),
$$
where $\nabla \psi=e_i \partial^i \psi$ and analogously for $H^{L}_X$.

Suppose subgroups ${\cal H}^L$ and ${\cal H}^R$ of $G$ with Lie algebras
$h^L$ and $h^R$ act on $T^{*}G$ by left and right translations
respectively. As it was mentioned above these actions are hamiltonian
and the corresponding momenta maps
$\mu^L:~T^{*}G\rightarrow (h^L)^*$ and $\mu^R:~T^{*}G\rightarrow (h^R)^*$
are given by
\begin{equation}
\begin{array}{ll}
<\mu^L(g,\xi),X>=<\xi,g^{-1}Xg>,&X\in h^L,  \\
<\mu^R(g,\xi),X>=-<\xi,X>,      &X\in h^R.
\end{array}
\label{mom}
\end{equation}

Consider the reduction of $T^{*}G$ under the action of
${\cal H}=({\cal H}^L, {\cal H}^R)$.
In accordance with the general Hamiltonian reduction procedure
\cite{Arn} the quotient space
$$
{\cal P}_{\mu}={\mu}^{-1}(T^{*}G)/\mbox{Stab}{~\mu}
$$
of the constant moment surface $\mu=const$ under its stabilizer
$\mbox{Stab}{~\mu}$ in ${\cal H}$ can be supplied under some additional
conditions by the structure of the symplectic manifold. The Poisson
bracket $\{,\}_{{\cal P}_{\mu}}$ on the reduced space can be found
by using the set $\mbox{Fun}(T^{*}G)^{\cal H}$ of $\cal H$-invariant
functions on $T^{*}G$.
Namely,
\begin{equation}
\{\phi,\psi\}_{{\cal P}_{\mu}}=\{\phi,\psi\}|_{\mu=const},~~~\phi,\psi\in
\mbox{Fun}(T^{*}G)^{\cal H}.
\label{red}
\end{equation}

Now let $G$ be a group of real (twisted) loops and
$G=N_+A K$ be its Iwasawa decomposition described above.
Put ${\cal H}^L=N_+$ and ${\cal H}^R=K$.
Following an analogy with the finite dimensional situation \cite{FO,ABT},
choose $\mu^R=0$ and $\mu^L=I=\sum_{i=o}^l f_i\in n_-=(n_+)^*$.

\begin{lemma}
Stabilizer of $I$ in $n_+^*$ coincides with $N_+$.
\end{lemma}
Proof. The Lie algebra Stab $I$ is determined by the condition
\begin{equation}
<I,[X,Y]>=0,~~~X,Y\in {\cal L}ie(\mbox{Stab}~I),
\label{cj}
\end{equation}
where $<,>$ stands for a bilinear symmetric form on $\cal G$, which is
compatible both with the homogeneous and with the principal gradation.
Let us show that if $X,Y \in n_+$, then (\ref{cj}) is satisfied.

Let us choose on ${\cal G}$ the homogeneous gradation.
For any homogeneous elements $X,Y\in n_+$ one has $\deg{X},\deg{Y}\geq 0$,
so that $\deg{[X,Y]}\geq 0$. Since $I\in {\cal G}^{-1}\oplus {\cal G}^0$
and $<,>$ is compatible with the homogeneous gradation:
$<{\cal G}^{k},{\cal G}^{m}>=\delta_{k,-m}$, only two nontrivial cases
should be considered:\\
a) $[X,Y]\in \bar{n}_+$, i.e. the commutator is generated by elements
$e_i$, $1\leq i\leq l$ and thereby does not contain any simple root in
its decomposition over the root basis in $\bar{\cal G}$.
Thus, we get (\ref{cj}).\\
b) $[X,Y]\in {\cal G}^{1}$, i.e. one of elements, say $X$, includes
one generator $e_0$. Then $<f_0,[X,Y]>=-<[X,f_0],Y>$. By taking into account
that $[X,f_0]\in \bar{n}_+$ and by using the isotropy of $\bar{n}_+$
we arrive at (\ref{cj}).\\

The case of the principle gradation is treated analogously.

It follows from Proposition 3 that the stabilizer of the chosen momenta
coincides with the group $N_+\times K$. With the help of this symmetry group,
any point $(g,\xi)\in T^{*}G$ lying on the surface of the fixed value
of the momenta map can be brought to the form
$(Q,L)$, where $Q\in A$ in the Iwasawa decomposition.
The pair $(Q,L)$ defines the coordinates on the reduced space
${\cal P}_{\mu}$. The explicit form of $L$ is fixed
by requiring $(Q,L)$ to be on the surface $\mu^R=0$, $\mu^L=I$:
\begin{equation} <L,X>=0~~\mbox{for}~~\forall X\in {\cal K},~~
<Q^{-1}LQ,X>=<I,X>~~\mbox{for}~~\forall X\in n_+.
\label{ww}
\end{equation}
Since ${\cal K}$ and $\cal M$ are orthogonal the first equation gives
\begin{equation}
L^{hom}=\sum_{\stackrel{k\geq 0}{1\leq i\leq l}}a_i^k(h_i^k+h_i^{-k})+
\sum_{\stackrel{\alpha\in \Delta}{k\geq 0}}
a_{\alpha}^{k}(E_{\alpha}^{k}+E_{-\alpha}^{-k}),
\label{st}
\end{equation}
\begin{equation}
L^{prin}=\sum_{\stackrel{k\geq 0}{1\leq i\leq l}}a_i^k(h_i^{kh}+h_i^{-kh})+
\sum_{\stackrel{k>0,k\neq 0 \bmod h}{\alpha\in \Delta^{k \bmod h}}}
a_{\alpha}^{k}(E_{\alpha}^{k}+E_{-\alpha}^{-k}).
\label{can}
\end{equation}
Here $L^{hom}$ and $L^{prin}$ denote $L$ in the homogeneous and in the
principal gradation respectively. Since $A$
in the Iwasawa decomposition coincides with the group of {\it constant}
loops being the exponent of $\bar{h}$, then $Q=e^q$, $q\in
\bar{h}$.  Equation (\ref{ww}) can be rewritten as
$P_{n_-}(e^{-q}Le^q)=I$, where $P_{n_-}$ is the projector on $n_-$.
Substituting these $L$ in (\ref{ww}) and using the identity
$e^{-q}E_{\pm\alpha}^{k}e^q=e^{\mp \alpha(q)}E_{\pm\alpha}^{k}$,
we get in both cases one formal expression for $L$:
\begin{equation}
L=\sum_{1\leq i\leq l}\pi_i h_i+
\sum_{1\leq i\leq l}
e^{-\alpha_i(q)}(e_i+f_i).
\label{LL}
\end{equation}
The element $L$ appeared in such a way is nothing but the
$L$-operator of the periodic Toda lattice \cite{Bog}. Taking a
matrix representation for $\bar{\cal G}$ and substituting in (\ref{LL})
the homogeneous or principal realization of Chevalley generators
one obtains the $L$-operator of the periodic Toda lattice as a matrix
function of the spectral parameter $\lambda$.

\section{Construction of $r$-matrices for the periodic Toda lattice.}
In this section we show how one can calculate $r$-matrices of the periodic
Toda lattice by using the Poisson bracket on the reduced phase space.

We start with the finite-dimensional situation.
Let $g=nQk$ be the Iwasawa decomposition of an element from some
semi-simple Lie group.  Function $F_X(Q,L)$, $X\in {\cal M}$
defined on ${\cal P}_{\mu}$ has an extension $F_X(g,\xi)=<\xi,k^{-1}Xk>$ on
$T^{*}G$ that is invariant with respect to the action of ${\cal
H}=N_+\times K$. In \cite{ABT} it was proved that
\begin{lemma}
The Poisson bracket on the reduced space has the form
\begin{equation}
\{F_X,F_Y\}(Q,L)=<L,[r(X),Y]+[X,r(Y)]>,
\label{baz}
\end{equation}
where the linear operator $r:~{\cal M}\rightarrow {\cal K}$
is given by the formula
$$ r(X)=\frac{d}{dt}|_{t=0}k(ge^{tX}). $$
\end{lemma}
Let us remark that (\ref{baz}) follows immediately from (\ref{red}).

Denote by $X=X_++X_a+X_k$ the Iwasawa decomposition of $X\in {\cal M}$,
where $X_+\in n_+$, $X_a\in a$, $X_k\in {\cal K}$. It can be shown
\cite{ABT} that
\begin{equation}
r(X)=X_k.
\label{mat}
\end{equation}
Clearly, this formula admits a straightforward generalization to the
infinite-dimensional case.\\

1. $r$-matrix in the homogeneous gradation.\\
Substituting the decomposition of an arbitrary
$x$ over the homogeneous basis in (\ref{mat})
$$ r\left(\sum_{\stackrel{\alpha\in
\Delta}{n>0}}x_{\alpha}^{n}W_{\alpha}^{n} +\sum_{\alpha\in \Delta_+}
x_{\alpha}^{0}W_{\alpha}^{0}+ \sum_{\stackrel{1\leq i\leq l}{n\geq 0}}
x_i^n T_i^n\right)= -\sum_{\stackrel{\alpha\in
\Delta}{n>0}}x_{\alpha}^{n}V_{\alpha}^{n} +\sum_{\alpha\in \Delta_+}
x_{\alpha}^{0}V_{\alpha}^{0}+ \sum_{\stackrel{1\leq i\leq l}{n>0}} x_i^n
H_i^n $$
and using the orthogonal relations for the basis elements
\begin{equation}
\begin{array}{l}
<T_i^n,T_j^m>=2(h_i,h_j)\delta_{n,m},~~~n,m>0, \\
<W_{\alpha}^n,W_{\beta}^m>=2(e_{\alpha},e_{-\alpha})
\delta_{\alpha\beta}\delta_{n,m},~~~n,m>0,
\end{array}
\end{equation}
we find the explicit form of the operator $r$ (-1/2 is omitted):
\begin{equation}
r=
\sum_{\stackrel{k>0}{1\leq i,j\leq l}}a^{ij}H_i^k <T_j^k,\ldots >+
\sum_{\alpha\in \Delta_+}V_{\alpha}^0<W_{\alpha}^0,\ldots >+
\sum_{\stackrel{k>0}{\alpha\in \Delta}}V_{\alpha}^k<W_{\alpha}^k,\ldots >,
\label{rm}
\end{equation}
where $a^{ij}$ is the inverse  matrix  to $(h_i,h_j)$, $1\leq i,j\leq l$.
This operator defines the element of the space
${\cal K}\otimes {\cal M}$ that we also denote by $r$:
\begin{equation}
r=
\sum_{\stackrel{k>0}{1\leq i,j\leq l}}a^{ij}H_i^k \otimes T_j^k+
\sum_{\alpha\in \Delta_+}V_{\alpha}^0\otimes W_{\alpha}^0 +
\sum_{\stackrel{k>0}{\alpha\in \Delta}}V_{\alpha}^k\otimes
W_{\alpha}^k.
\label{rm1}
\end{equation}
Nota that as an element of ${\cal K}\otimes {\cal M}$
the $r$-matrix is not uniquely defined.
Since in (\ref{baz}) elements $L,X,Y\in {\cal M}$,
we can always add to (\ref{rm1}) an arbitrary vector from
$ {\cal M}\otimes {\cal K}$ without changing the value of the bracket.
Thus, we define $R=r-\sigma(r)$ where $\sigma$
is the permutation operator in ${\cal G}\otimes{\cal G}$. Substituting in
$R$ the explicit form (\ref{lb}) of the generators, we get
\begin{equation}
R^{hom}=
\sum_{\stackrel{k>0}{1\leq i,j\leq l}}a^{ij}(h_i^k \otimes h_j^{-k}
-h_i^{-k} \otimes h_j^{k})+
\sum_{\alpha\in \Delta_+}(e_{\alpha}^{0}\otimes e_{-\alpha}^0 -
e_{-\alpha}^{0}\otimes e_{\alpha}^0)+
\label{rmR}
\end{equation}
$$
\sum_{\stackrel{k>0}{\alpha\in \Delta}}
(e_{\alpha}^k\otimes e_{-\alpha}^{-k}-e_{-\alpha}^{-k}\otimes
e_{\alpha}^{k}),
$$
where $R^{hom}$ stands for the $r$-matrix in the homogeneous
gradation.  $R^{hom}\in {\cal G}\otimes{\cal G}$
constructed in such a way coincides with the classical trigonometric
$r$-matrix of the periodic Toda lattice
corresponding to the homogeneous gradation of $L$-operator (\ref{LL}).
Namely, identifying ${\cal G}\otimes{\cal G}$ with ${\cal
G}[\lambda,\lambda^{-1}]\otimes{\cal G}[\mu,\mu^{-1}]$, i.e.  substituting
in (\ref{rmR}) the homogeneous realization of the generators and performing
the summation we find
\begin{equation}
R^{hom}=\frac{\lambda r_--\mu r_+}{\lambda-\mu},
\label{SDJ}
\end{equation}
where $r_+=\frac{1}{2}a^{ij}h_i\otimes h_j+\sum_{\alpha\in \Delta_+}
e_{\alpha}\otimes e_{-\alpha}$, $r_{-}=-\sigma (r^+)$ are
two constant solutions of the classical Yang-Baxter equation.

2. $r$-matrix in the principle gradation.\\
The case of the principle gradation is completely analogous to the
previous one.
We write down the explicit form of $r$ and $R^{prin}$ that are
the corresponding $r$-matrices in the principle gradation:
\begin{equation}
r=
\sum_{\stackrel{k>0}{1\leq i,j\leq l}}a^{ij}H_i^{kh} \otimes T_j^{kh}+
\sum_{\stackrel{k>0}{k\neq 0 \bmod h}}
\sum_{\alpha\in \Delta^{k\bmod h}}V_{\alpha}^k\otimes
W_{\alpha}^k,
\label{rm2}
\end{equation}
\begin{equation}
R^{prin}=
\sum_{\stackrel{k>0}{1\leq i,j\leq l}}a^{ij}(h_i^{kh} \otimes h_j^{-kh}
-h_i^{-kh} \otimes h_j^{kh})+
\label{rmR1}
\end{equation}
$$
\sum_{\stackrel{k>0}{k\neq 0 \bmod h}}\sum_{\alpha\in \Delta^{k\bmod h}}
(e_{\alpha}^k\otimes e_{-\alpha}^{-k}-e_{-\alpha}^{-k}\otimes
e_{\alpha}^{k}).
$$
Performing the summation in the last formula, we get
the well-known solution \cite{BD} of the classical Yang-Baxter
equation corresponding to the principle gradation of $\cal G$:
\begin{equation}
R^{prin}=\frac{t_0}{2}+\frac{1}{x^h-1}\sum_{k=0}^{k=h-1}t_kx^k,
\end{equation}
where $x=\frac{\lambda}{\mu}$, $t_0=a^{ij}h_i\otimes h_j$ and
$t_k=\sum_{\alpha\in \Delta^{k}}e_{\alpha}\otimes e_{-\alpha}$.

In conclusion, the method of the Hamiltonian reduction applied
to the cotangent bundle over a loop group allows one to construct
$r$-matrices depending on the spectral parameter. Let us stress that
the possibility to perform the reduction considered in the paper
is due to the existence of the Iwasawa decomposition
$G=N_+AK$, in which the group $A$ turns out to be finite-dimensional, so
that the reduced phase space is also finite-dimensional. The situation
described above fails for the Cartan decomposition
$G=KAK$ of a complex loop group $G$, i.e. the factor $A$ this time is
infinite-dimensional. Note that in the finite-dimensional setting
the Cartan decomposition is needed to perform the Hamiltonian reduction
$K\backslash T^*G/ K$ leading to
Calodgero-Moser-Sutherland models \cite{ABT}. Seemingly,
$r$-matrices with the spectral parameter for these models can be obtained by
the Hamiltonian reduction of $T^*G(\lambda)$ or $T^*\hat{{\cal G}}(\lambda)$
\cite{GN}, where instead of ${\cal H}= K\times K$ some other
group of diffeomorphisms of a loop group or a loop algebra should be used.
We hope to study this question in further publications.
$$~$$
{\bf ACKNOWLEDGMENT}
$$~$$ The author is grateful to A.Gorsky, P.Medvedev and
S.Frolov for useful discussions. This work is supported in part by RFFR
under grant N93-011-147 and by ISF under grant M1L-000.

\end{document}